\documentclass[final,1p,times]{elsarticle}

\usepackage{graphicx}

\usepackage{amssymb}
\usepackage{amsmath}
\usepackage{siunitx}
\usepackage{caption}
\usepackage{subcaption}

\journal{Experimental Thermal and Fluid Science}

\begin{document}

\begin{frontmatter}

\title{Optimized thermoelectric sensitivity measurement for differential thermometry with thermopiles}

\author{Tim Prangemeier\corref{cor1}\fnref{fcor1}}
\fntext[fcor1]{email: prangemeier@bcs.tu-darmstadt.de}

\author{Iman Nejati\corref{cor3}\fnref{fcor3}}
\fntext[fcor3]{email: nejati@ttd.tu-darmstadt.de}

\author{Andreas M\"{u}ller, \\Philip Endres, Mario Fratzl\corref{cor4}}
\author{Mathias Dietzel\corref{cor2}\fnref{fcor2}}
\cortext[cor2]{Tel.: +49 6151 16-24276}
\fntext[fcor2]{email: dietzel@nmf.tu-darmstadt.de}

\address{Institute for Nano- and Microfluidics,\\
Center of Smart Interfaces, TU Darmstadt,\\
 Alarich-Weiss-Str. 10, 64287 Darmstadt, Germany}

\begin{abstract}

A novel approach to calibrate the sensitivity of a differential thermometer, consisting of several thermocouples connected in series (thermopile), has been developed. The goal of this method is to increase the accuracy of small temperature difference measurements ($\Delta T \leq \SI{1}{\kelvin}$), without invoking higher sensor complexity. To this end, a method to determine the optimal temperature difference employed during the differential measurement of thermoelectric sensitivities has been developed. This calibration temperature difference is found at the minimum of combined measurement and linearization error for a given mean temperature. The developed procedure is demonstrated in an illustrative example calibration of a nine-junction thermopile. For mean temperatures between $\SI{-10}{\celsius}$ and $\SI{+15}{\celsius}$, the thermoelectric sensitivity was measured with an uncertainty of less than $\SI{\pm 2}{\percent}$. Subsequently, temperature differences as low as $\SI{0.01}{\kelvin}$ can be resolved, while the thermometer used for the example calibration was accurate only to $\SI{\pm0.3}{\kelvin}$. This and higher degrees of accuracy are required in certain research applications, for example to detect heat flux modulations in bifurcating fluidic systems.

\end{abstract}

\begin{keyword}
thermopile calibration \sep thermoelectric sensitivity measurement \sep Seebeck coefficient measurement \sep Seebeck effect \sep differential thermometer \sep calibration uncertainty analysis \sep high-precision heat flux modulation measurement

\indent{\rule{13.7cm}{0.4pt}} 
This is the post-print authors' version of the manuscript published in Experimental Thermal and Fluid Science. \copyright \ Elsevier 2015. doi: 10.1016/j.expthermflusci.2015.01.018. \\ This manuscript version is made available under the CC-BY-NC-ND 4.0 license http://creativecommons.org/licenses/by-nc-nd/4.0/.


\end{keyword}

\end{frontmatter}

\section{Introduction}
\label{sec:intro}

In 1826 A. C. Becquerel made the first recorded proposal to employ the Seebeck effect in thermometers and by 1902 thermocouples were commercially available \cite{Hunt1964}. Today, they are in widespread use due to their low cost, robustness, small size, simplicity, speed of response and large temperature range \cite{Childs2000}. This thermoelectric effect was discovered in the early 1820s by T. Seebeck \cite{vanHerwaarden1986,Bell2008,DiSalvo1999}, whereby an electric potential is induced due to a temperature gradient in a thermopair \cite{Riffat2003,Huang1990,Martin2010}. It is often utilized in specialized devices, such as thermometry near absolute zero \cite{Armbruester1981,Maeno1983} or in nanoscale devices \cite{Bakker2012}. 

While the absolute temperature is of interest in many applications, in others, such as monitoring nuclear reactors \cite{Hashemian2006}, temperature difference measurements are required. For example, in the 1970s thermocouples were calibrated to measure temperature differences in aircraft engine oil as low as $\SI{2.5}{\kelvin}$ over an absolute temperature range of hundreds of Kelvin \cite{Eccles1973}.

Thermocouples can be employed to directly determine a temperature difference in a single measurement (and without reference junction compensation) \cite{Huang1990,Martin2010}. This circumvents the error propagation that would otherwise be encountered when differencing two separate temperature measurements. To increase sensitivity, multiple thermocouples can be connected in series forming a thermopile. For precise measurements of differential temperatures above $\SI{1}{\kelvin}$, \citet{Huang1990} calibrated a thermopile with a high precision quartz thermometer $(T \pm \SI{0.04}{\kelvin})$ and a specialized method of signal conversion. An accuracy of $\pm \SI{0.07}{\kelvin}$ was achieved. When the calibrated thermopile is employed to measure differential temperatures, the mean temperature $T_{\text{\text{m}}}$ is used to account for the non-linearity of the calibration curve. This method is limited by the accuracy of the calibration thermometer \cite{Drnovsek1998}.

Differential temperature measurements are of particular interest for heat transfer investigations, as heat flux cannot be measured directly \cite{Childs1999}. Nonetheless, differential thermometry can be used to relate the heat flux to the temperature gradient and the material properties (known heat resistance). Heat flux sensors based on this technique have been designed for a variety of applications ranging from industry to biological systems research, as well as radiometry for photo-voltaic and solar thermal energy studies \cite{Langley1999, Ewing2010, Baughn1986, Park2012}. Heat flux uncertainty of $\SI{\pm 4}{\percent}$ and $\SI{\pm 7}{\percent}$ are reported by \cite{Baughn1986} and \cite{Park2012} respectively.

In many heat transfer applications, the heat resistance between two thermal reservoirs is not known a priori. In these cases, the heat flux can be determined by measuring the heating (or cooling) power required to maintain quasi-isothermal reservoir boundaries. For electric heat sources this is simply achieved by an electric-power measurement. For connectively cooled heat sink surfaces, on the other hand, this is typically accomplished by measuring the difference between the inlet and outlet temperature of the coolant in conjunction with its mass flux and material properties \cite{Koschmieder1974b}. A high degree of temperature uniformity of the cooled plate is required in many studies. The pattern symmetry of surface tension gradient driven B\'enard-Marangoni convection, for example, is highly sensitive towards temperature non-uniformities. Therefore, accurate differential temperature measurements are required to detect the heat flux modulation caused by bifurcations points, while at the same time maintaining a thermal gradient across the plate, which is as small as possible in order to maintain quasi-isothermal boundary conditions. To this end, a thermoelectric circuit has been developed by \cite{Koschmieder1974b}; however, no uncertainty propagation analysis has been reported for the sensor calibration. 

The thermoelectric sensitivity (TES) is employed in the above mentioned heat flux studies \cite{Langley1999, Koschmieder1974b, Ewing2010}. It is commonly measured with the differential method (described in \cite{Martin2010}) \cite{Ewing2010,Wold1916, Bidwell1922,Weiss1956, Burkov2001, Zhou2005}. The uncertainty of the achieved TES measurement is dependent on the chosen temperature difference. To avoid non-linear effects and errors, it has in the past been recommended to choose a temperature difference on the order of a few percent of the mean temperature $T_{\text{\text{m}}}$ ($\Delta T/ T_{\text{\text{m}}} << 1$) \cite{Zhou2005,Martin2010}. However, until now, there has been no report of any systematic study of the selection of the temperature difference employed during TES measurement. This shortcoming is addressed in this study.

While international standards for the calibration of thermocouples as thermometers exist, customized calibration techniques have been developed for specialized applications \cite{Huang1990,Ballestrin2006b}. For example, calorimetric and radiometric calibrations of thermoelectric heat flux sensors and radiometers have been developed \cite{Ballestrin2004, Ballestrin2006b}. For precise measurements of small differential temperatures, a method of calibration and signal conversion has been proposed for measurements with thermopiles \cite{Huang1990}. In this study we address the calibration of even smaller temperature differences, and propose a novel approach.

Subsequently, the above mentioned technique proposed by \citet{Huang1990} is further extended to measure temperature differences below $\SI{1}{\kelvin}$. In addition to the mean temperature of the differential temperature measurement (taken into account by \citet{Huang1990}), the mean temperature during the calibration process is considered as well. Instead of the calibration of $U(T)$, the inverse of the TES is used as the calibration coefficient, and it is measured with the differential method described in \citet{Martin2010}. An analysis of the maximum uncertainty of this method is performed and a procedure to optimize the accuracy of the differential technique is developed. The novel method is employed over a range of temperatures ($\SI{-10}{\celsius}$ to $\SI{15}{\celsius}$) in an illustrative example and compared to the aforementioned technique. A higher differential temperature resolution is achieved, although a significantly less accurate calibration thermometer $(T \pm \SI{0.3}{\kelvin})$ was used. Finally, the steps of this method from calibration to small differential temperature measurement with high accuracy are summarized.

\begin{table}[h!]
\centering
\begin{tabular}{p{15mm}p{110mm}} 
$\mathrm{\textbf{Nomenclature}}$ & \\
\hline \hline\\[1mm]
$2h $   & applied calibration temperature difference ($\si{\kelvin}$)\\[0mm]
$2h_{\text{opt}}$ & optimal temperature difference ($ \si{\kelvin} $) \\
$E_{\text{\text{m}}} $   & measurement uncertainty ($\si[per-mode = fraction]{ \volt \per \kelvin}$)\\[0mm]
$E_{\text{lin}} $ &  linearization error ($\si[per-mode = fraction]{\volt \per \kelvin}$)\\[0mm]
$E_S $   &  uncertainty of TES ($\si[per-mode = fraction]{\volt \per \kelvin}$)\\[0mm]
$E_{T,\text{cal}} $   & temperature uncertainty of calibration thermometer ($ \si{\kelvin} $)\\[0mm]
$S_{\text{\text{A}}},S_{\text{\text{B}}}$ & TES of materials $\text{\text{A}}$, $\text{\text{B}}$ ($\si[per-mode = fraction]{ \volt \per \kelvin}$)\\
$S_{\text{\text{AB}}}$ & TES of material pairing $\text{\text{AB}}$ ($\si[per-mode = fraction]{\volt \per \kelvin}$)\\
$S_{\text{st}} $ & TES from industry standard ($\si[per-mode = fraction]{\volt \per \kelvin}$)\\[0mm]
$S_{\text{st,K}} $ & standard K-type TES ($\si[per-mode = fraction]{\volt \per \kelvin}$)\\[0mm]
$S(T) $   & TES as a function of temperature ($\si[per-mode = fraction]{\volt \per \kelvin}$)\\[0mm]
$T$ & temperature ($\si{\celsius}$)\\[0mm]
$ T_{\text{\text{m}}} $  & mean temperature ($\si{\celsius}$)\\[0mm]
$ T_{\text{ref}}$   &reference temperature ($\si{\celsius}$)\\[0mm]
$\Delta T $    & temperature difference ($\si{\kelvin}$)\\[0mm]
$U(T,T_{\text{ref}}) $   &electric potential difference across thermopile ($\si{ \volt}$)\\[0mm]
$U(T_1,T_2) $   &electric potential difference across thermopile ($\si{ \volt}$)\\[0mm]

$ $  &\\[0mm]

\hline\\[-1mm]
\end{tabular}
 \label{tab:nomenclature}
\end{table}

\section{Fundamentals of thermocouples and thermopiles}
\label{sec:theory}

\begin{figure}[h]
\centering
{\includegraphics[]{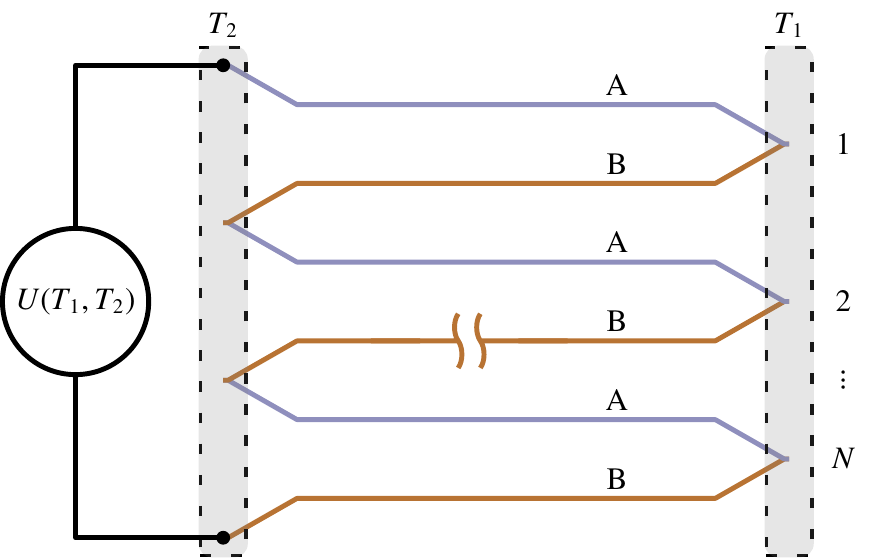}}
\caption{Schematic representation of a thermopile, composed of $N$ thermopairs of materials $\text{\text{A}}$ and $\text{\text{B}}$. The temperature difference between $T_1$ and $T_2$ leads to the electrical potential difference $U(T_1,T_2)$.
}
\label{fig:thermocouple}
\end{figure}

The TES or Seebeck coefficient $S_{\text{A}}(T)$ is a physical property of material A and is dependent on the local temperature $T$ \cite{Martin2010,Huang1990}. As expressed by the Thomson relations (which essentially express microscopic reversibility), the Seebeck effect itself is a manifestation of the cross-correlation between thermal transport due to a gradient in electric potential on the one hand and charge separation induced by temperature differences on the other. Hence, when the material is exposed to an infinitesimal temperature difference $\mathrm{d}T$, an electric potential difference $\mathrm{d}U$ is induced:
 \begin{equation}\label{eqn:dU}
\mathrm{d}U = S_{\text{A}}(T) \mathrm{d}T \mathrm{.}
\end{equation}
The TES of a thermocouple is the difference between the sensitivities of materials A and B ($S_{\text{\text{AB}}} = S_A - S_B$). Thus, for a thermocouple ($N=1$) exposed to a finite temperature difference $T_2-T_1$, an electric potential difference 
 \begin{equation}\label{eqn:vt}
U(T_{1},T_{2}) = N \cdot \int_{T_1}^{T_2}{S_{\text{\text{AB}}}(T) \mathrm{d}T}\mathrm{,}
\end{equation}
is induced \cite{Martin2010,Bernhard2004}. In order to improve the signal-to-noise ratio, the potential difference can be augmented by aligning multiple ($N>1$) thermocouples in series. This assembly is typically referred to as a thermopile \cite{Huang1990} and is shown schematically in Fig. \ref{fig:thermocouple}.

\begin{figure}
\centering
{\includegraphics[]{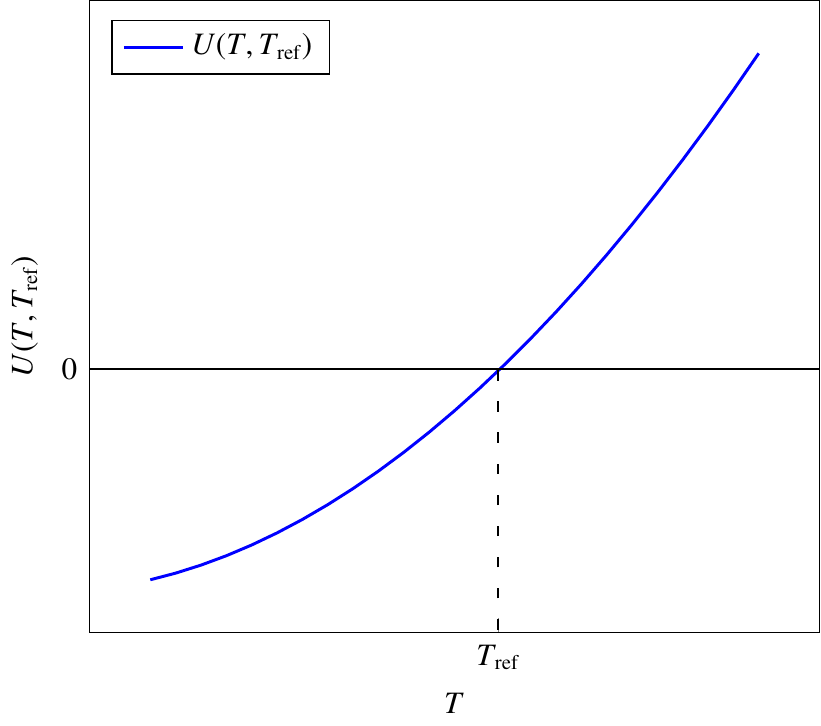}}
\caption{Schematic depiction of the potential difference $U(T,T_{\text{ref}})$ as a function of the measurement junction temperature $T$ and for a given reference temperature $T_{\text{ref}}$.
}
\label{fig:ut}
\end{figure}

In accordance with technical standards (\citet{DIN60584}, \citet{EURAMET}) thermopiles are commonly calibrated by comparison with a calibration thermometer. For later reference, we refer to this type of calibration as absolute calibration. One set of junctions is held at a constant reference temperature $T_{\text{ref}}$, while the other set of junctions is exposed to a range of $P$ different temperatures $T_i$ ($i=1,..,P$). The resulting calibration curve (with interpolated values between the $T_i$) is the potential difference $U(T,T_{\text{ref}})$ as a function of the temperature $T$ at the measurement junction. By definition $U(T_{\text{ref}},T_{\text{ref}})$ is zero at $T_{\text{ref}}$. Apart from this offset, the calibration curve is otherwise independent of $T_{\text{ref}}$ \cite{Huang1990,Bernhard2004,Bentley1998}.  In general the calibration relationship is a non-linear function of the temperature $T$. For various standard type thermocouples (E, J, K, N and R) \citet{Drebuschak2009} found $U(T,T_{\text{ref}})$ to be a linear function of $T$ at high temperatures and a quadratic function of $T$ at lower temperatures. Typically, the transition from the quadratic to the linear relation takes place between $\SI{250}{\kelvin}$ and $\SI{350}{\kelvin}$. In this study $U(T,T_{\text{ref}})$ is considered to increase monotonously with temperature, as is shown schematically in Fig. \ref{fig:ut}.

For the absolute calibration type, \citet{Drnovsek1998} analyzed the various sources of uncertainty and found the error $E_{T,\text{cal}}$ associated with the accuracy of the thermometer used for calibration to be the limiting factor. Therefore, measurements that make use of this type of calibration have an uncertainty of at least $E_{\Delta T} \geq \sum E_{T,\text{cal}}$. The sum of calibration thermometry uncertainty stems from the measurement of $T_{\text{ref}}$ and $T_i$.

\section{Theoretical derivation of optimized TES measurement method}
\label{sec:sens}

In order to circumvent the uncertainty limitations imposed by the absolute calibration, the TES of the thermopile  \mbox{$S(T) = N \cdot S_{\text{\text{AB}}}(T)$} is employed as the calibration parameter in this study. For small differential temperatures, Eq. (\ref{eqn:vt}) is linearized and the calibration relationship becomes:
\begin{equation}\label{eqn:linvt}
U(\Delta T,T_{\text{\text{m}}}) = {S(T_{\text{\text{m}}})} \cdot \Delta T \mathrm{.}
\end{equation}
As is detailed below, this correlation is relatively robust towards uncertainty in the measurement of the mean temperature $T_{\text{\text{m}}}$. The TES of a K-type thermocouple is listed in Table \ref{tab:st-20} (derived from \citet{DIN60584}). In the temperature range of interest in this study, the largest variation of TES due to a fluctuation of $T_{\text{\text{m}}}$ by $\SI{1}{ \kelvin}$ is found to be less than $\SI{0.2}{\percent}$ (between $\SI{-10}{ \celsius}$ and $\SI{-9}{ \celsius}$).

\begin{table}[h]
\centering
\caption{TES $S(T)$ for selected temperatures based on the industry standard \citet{DIN60584}.}
 \label{tab:st-20}
\begin{tabular}{c | c}
{temperature}&{TES }  \\[1mm]
{ $T$ in \si{\celsius}} & {$S(T)$ in \si[per-mode = fraction]{\micro \volt \per \kelvin}}\\
\hline
$-10 $  & $38.896$ \\[0mm]
$-9$ & $38.957$ \\[1mm]
$-8$ & $39.018$\\[1mm]
$-7$ & $39.077$\\[1mm]
$-6$ & $39.135$\\[1mm]
\hline
\end{tabular}
\end{table}

The TES $S(T_{\text{\text{m}}})$ of a thermopile can be experimentally determined by either the integral or the differential method \cite{Martin2010}. The integral (or large $\Delta T$) method involves differentiating an analytic expression of the calibration curve $U(T,T_{\text{ref}})$ \cite{Wood1988}. One drawback of this method is that small fluctuations of the reference temperature influence the accuracy of the absolute calibration curve $U(T,T_{\text{ref}})$. Furthermore, no objective method to evaluate the accuracy of the sensitivity exists \cite{Martin2010}.

For the differential (small $\Delta T$) method, a temperature difference of $2h$ is applied around $T_{\text{\text{m}}}$ and the resulting potential difference is measured \cite{Pinisetty2011}.  The sensitivity is then calculated as
\begin{equation}\label{eqn:stm}
S(T_{\text{\text{m}}}) = \frac{U(T_{\text{\text{m}}} + h)-U(T_{\text{\text{m}}}-h)}{2h} \mathrm{,}
\end{equation}
where the numerator is the voltage $U(T_{\text{\text{m}}} + h,T_{\text{\text{m}}}-h)$ measured across the thermopile. Expression (\ref{eqn:stm}) is effectively a second order central differencing scheme to calculate the gradient $U' (T_{\text{\text{m}}}) = S(T_{\text{\text{m}}})$ at $T_{\text{\text{m}}}$. This technique was already employed in 1916 by \citet{Wold1916} and is still employed today \cite{Middleton1953,Weiss1956,Burkov2001,Zhou2005,Pinisetty2011}. Typical temperature differences $2h$ range from $\SI{1}{\kelvin}$ to $\SI{20}{\kelvin}$. While most studies offer no analysis of measurement uncertainty, \citet{Burkov2001} achieved TES uncertainties on the order of $\pm \SI{4}{\percent}$ to $\pm \SI{10}{\percent}$. However, to the best of our knowledge, no study exists, which addresses the selection strategy of $2h$ itself.

Given the diametrical effect of the temperature difference $2h$ on the measurement uncertainty and linearization error, a procedure to select an optimal temperature difference $2h$ for the calibration is described in the following. To that end, the measurement uncertainty and linearization error are analyzed and an optimal value $2h$ is found, for which the combined error is minimized. At the same time the linearization error is quantified, allowing it to be monitored and limited relative to the measurement uncertainty.

\subsection{Measurement uncertainty}
\label{sec:me}

\begin{figure}
\centering
{\includegraphics[]{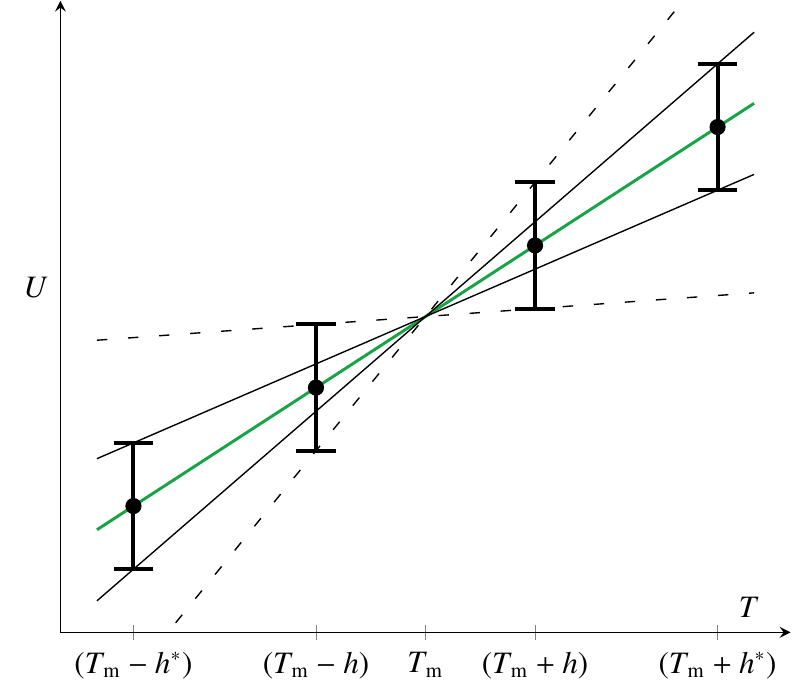}}
\caption{Schematic representation of TES measurement uncertainty variation, with a constant voltage measurement error denoted by the error bars. A larger uncertainty in the measured gradient results from the smaller $h$ (dashed) than for the larger $h^*$ (solid).}
\label{fig:me}
\end{figure}

When experimentally determining the TES $S(T_{\text{\text{m}}})$ with Eq. (\ref{eqn:stm}), the uncertainty of the temperature and voltage measurements is propagated. The maximum relative measurement uncertainty is given by:
\begin{equation}\label{eqn:me1}
\frac{E_{\text{m}}}{S(T_{\text{\text{m}}})} = \frac{E_{2h}}{2h} + \frac{E_U}{U(2h,T_{\text{\text{m}}})} \mathrm{,}
\end{equation}
where $E_{2h}=2E_{T,\text{cal}}$ and $E_{U}$ are the measurement uncertainty of the applied temperature difference $2h$ and of the potential difference $U$, respectively. Herein, the voltage $U(2h,T_{\text{\text{m}}})$ is approximated by $U(2h,T_{\text{\text{m}}}) = 2h \cdot N\cdot S_{\text{st,K}}(T_{\text{\text{m}}})$, with values for $S _{\text{st,K}}(T_{\text{\text{m}}})$  taken from an industry standard e.g. \citet{DIN60584}. This leads to the measurement uncertainty of the sensitivity as a function of the applied (calibration) temperature difference $2h$:
\begin{equation}\label{eqn:me2}
E_{\text{m}}(h) = \frac{E_U+ E_{2h}\cdot N \cdot S_{\text{st,K}}(T_{\text{\text{m}}})}{2h} \mathrm{.}
\end{equation}
Hence, the influence of the measurement uncertainty can be minimized by applying large temperature differences $2h$ for the calibration. This is depicted schematically in Fig. \ref{fig:me}.

\subsection{Linearization error}
\label{sec:le}
Determination of $S(T_{\text{\text{m}}})$ according to Eq. (\ref{eqn:stm}) approximates the slope of $U(T)$ at $T=T_{\text{\text{m}}}$ by means of linearization with a central difference scheme. Given the generally non-linear character of $U(T)$, neglecting the higher-order terms of the differencing scheme, introduces a linearization error. Using two Taylor series developed around $T_{\text{\text{m}}}$  and evaluated at $+h$ and $-h$ \cite{Schafer2006}, one finds
\begin{equation}\label{eqn:le1}
U'(T_{\text{\text{m}}}) = \frac{U(T_{\text{\text{m}}} +h) - U(T_{\text{\text{m}}} - h)}{2h} - \frac{U'''(T_{\text{\text{m}}})h^2}{3!}-( ...)\mathrm{.}
\end{equation}
Consequently, the linearization error reads

\begin{equation}\label{eqn:le2}
E_{\text{lin}}(h)=\frac{U'''(T_{\text{\text{m}}})h^2}{3!}+ (...)\mathrm{ .}
\end{equation}
Therefore, in contrast to the measurement uncertainty, the linearization error is minimal for small values of  $2h$. The trade-off between the measurement and linearization error leads to an optimization problem. In the following, an optimal  calibration temperature difference $2h_{\text{opt}}$, with a minimum of combined uncertainty, is derived.

\subsection{Optimization of total thermoelectric sensitivity uncertainty}
\label{sec:em}
 The sum of the measurement uncertainty (Eq. (\ref{eqn:me2})) and linearization error (Eq. (\ref{eqn:le2})) is an expression of the total uncertainty $E_{S}(h) = E_{\text{m}}(h)+E_{\text{lin}}(h)$ of the TES $S$. The minimum of $E_S(h)$ determines $h_{\text{opt}}$, i.e. ${\mathrm{d}E_S(h_{\text{opt}})}/{\mathrm{d}h} \stackrel{!}{=} 0$.

For an approximation of $U(T,T_{\text{ref}})$ with a polynomial of 3rd order, $h_{\text{opt}}$ can be found analytically:
\begin{equation}\label{eqn:hopt}
h_{\text{opt}} = \sqrt[3]{3 \frac{E_U+ E_{2h}\cdot N \cdot S_{\text{st}}(T_{\text{\text{m}}})}{2U'''(T_{\text{\text{m}}})}}\mathrm{.}
\end{equation}
For higher-order polynomials $h_{\text{opt}}$ is computed numerically. 

\subsection{Sensitivity calibration curve}
\label{sec:sc}
The method described above yields the TES $S(T_{\text{\text{m}}})$ of a thermopile at a given mean temperature $T_{\text{\text{m}}}$ with minimized total uncertainty. Generally, thermopiles are calibrated  for measurements over a range of mean temperatures. Therefore, the aforementioned procedure is repeated at intervals over the temperature range of interest, leading to a sensitivity calibration curve $S(T)$ as a function of temperature. 

\section{Illustrative calibration of a thermopile}
\label{sec:illex}

The application of the above derived TES measurement method is demonstrated in the following illustrative example. The thermopile considered here is used for the detection of heat flux modulation in bifurcating fluidic systems. In this particular experimental setup, geometric boundary conditions limit the number of thermopile junctions and require the thermocouple wiring to be relatively long (over $\SI{0.5}{\meter}$). However, this method is applicable to thermopiles in general, and its effectiveness is expected to increase with the number of junctions (as later described in Section \ref{sec:disc}).

\subsection{Experimental setup and procedure}
\label{sec:exp}

\begin{figure}[]
\centering
\caption{Schematic of the experimental setup for the calibration of $U(T,T_{\text{ref}})$ and $S(T)$. }
{\includegraphics[]{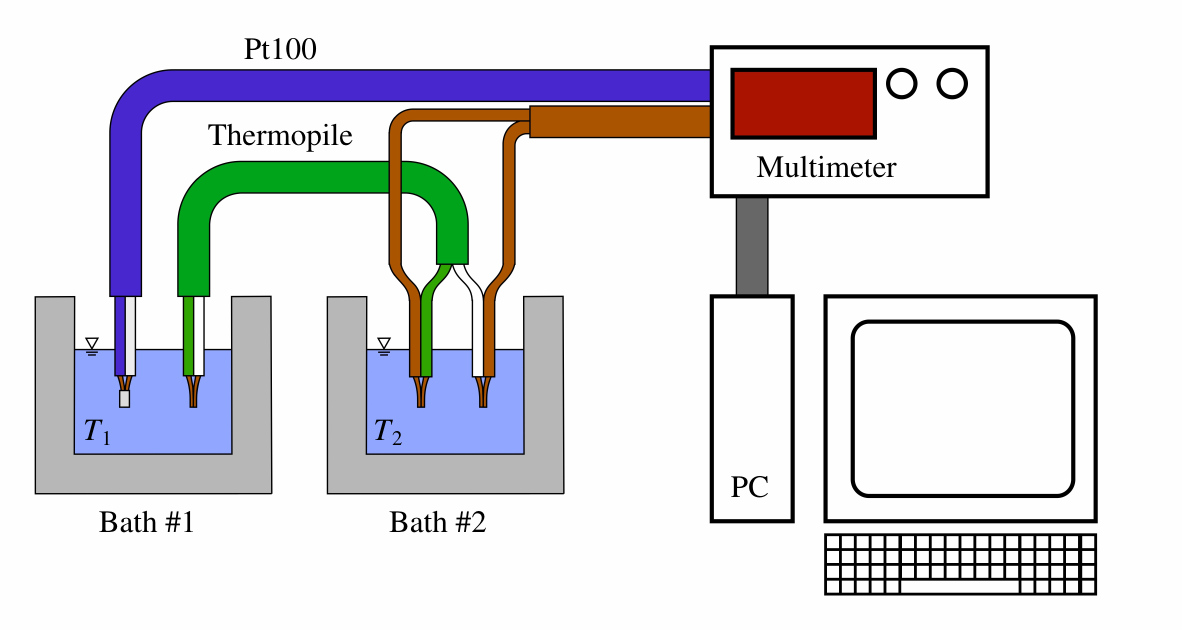}}
\label{fig:apparatus}
\end{figure}

The thermopile employed in this study was fabricated with $N = 9$ K-type thermocouples, each of $\SI{700}{mm}$ length, connected in series. Two calibration baths (Brookfield TC550 and Polyscience 1197P), with temperature stabilities of $\pm \SI{0.01}{\kelvin}$, were employed. A Pt100 resistance thermometer, with measurement uncertainty $E_{T,\text{cal}} = \pm \SI{0.3}{\kelvin}$, was utilized to measure the temperatures of the baths. The same thermometer was used for every measurement  in order to reduce the effect of a systematic thermometer offset. A Fluke 8846A multimeter was used to measure the potential difference across the thermopile with an uncertainty of $E_U = \pm \left( \SI{3.5}{\micro \volt} + \SI{0.0025}{\percent} \hspace{1mm}U \right)$. The apparatus is shown schematically in Fig. \ref{fig:apparatus}.

The calibration curve $U(T,T_{\text{ref}})$ was recorded at intervals of $\SI{5}{\kelvin}$ over a temperature range from $\SI{-30}{\celsius}$ to $\SI{40}{\celsius}$. The reference temperature was $T_{\text{ref}}= \SI{10}{\celsius}$. Based on this and with the procedure described in Section \ref{sec:sens}, the optimal TES calibration temperature difference $2 h_{\text{opt}}$ was calculated for each sensitivity measurement. The TES was then measured in $\SI{5}{\kelvin}$ intervals between $\SI{-10}{\celsius}$ and $\SI{15}{\celsius}$. In some cases limitations on the minimum and maximum temperatures of the calibration baths hindered the application of $2h_{\text{opt}}$. In these cases an alternate calibration temperature difference $2h$ was employed. Four repetitions of the TES measurements were performed for each temperature interval. 

Inhomogeneous thermocouple material in combination with temperature gradients can lead to voltage offsets \cite{Martin2010}, for which the calibration relation expressed in Eq. (\ref{eqn:linvt}) does not account for. However, no such offset was found when a strong temperature gradient was induced by exposing both ends of the thermopile to the same low temperature (approximately  $\SI{-30}{\celsius}$), while the middle section remained at room temperature. Therefore, such offsets are deemed to be negligible throughout the measurements. 

\subsection{Results}
\label{sec:results}
\begin{figure}
\centering
{\includegraphics[]{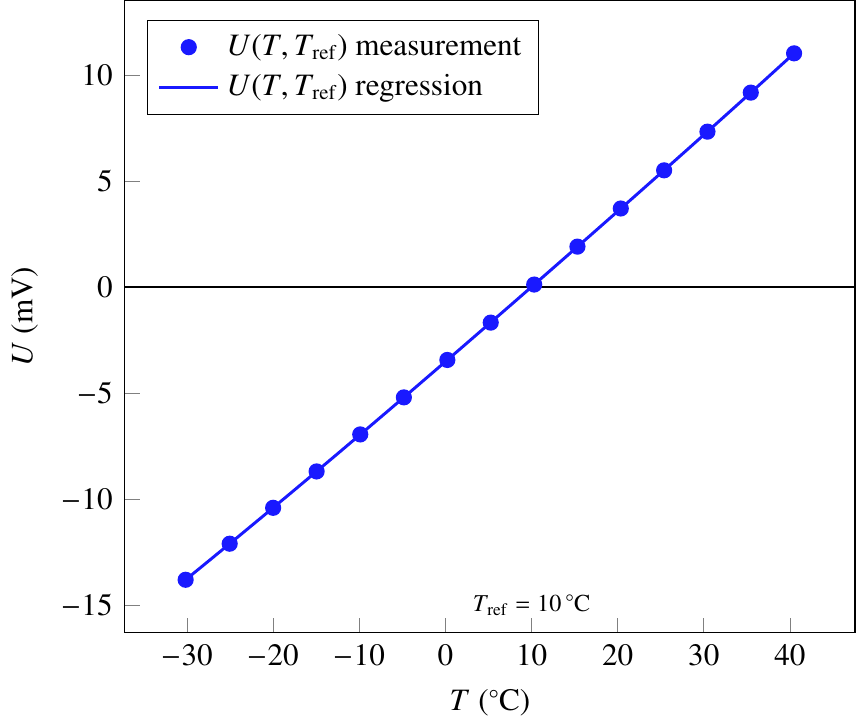}}
\caption{Absolute calibration measurement data and polynomial regression of $U(T,T_{\text{ref}})$.
}

\label{fig:exput}
\end{figure}

\begin{figure}
\centering
{\includegraphics[]{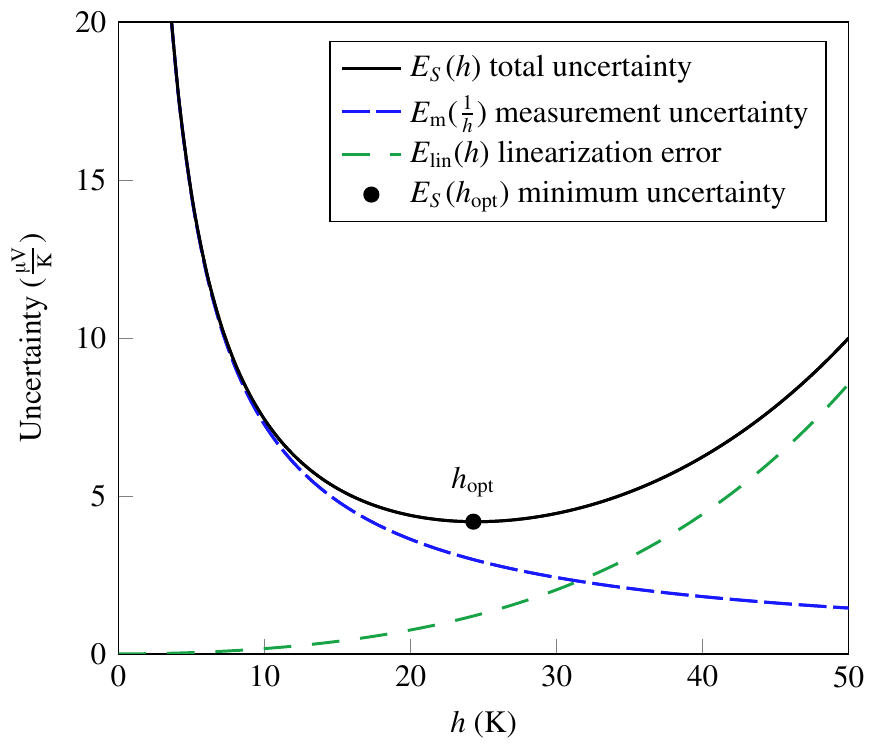}}
\caption{ The sensitivity measurement uncertainty $E_{\text{m}}(\frac{1}{h})$ (tightly dashed blue), linearization error $E_{\text{lin}}(h)$ (loosely dashed green) and the total TES calibration uncertainty $E_{S}(h)$ (solid black), for a sensitivity measurement at $T_{\text{\text{m}}} = \SI{0}{\celsius}$. The minimum combined error is found at $h_{\text{opt}}$. 
}
\label{fig:pler}
\end{figure}

The recorded calibration curve $U(T,T_{\text{ref}})$ is shown in Fig. \ref{fig:exput}. A  5th order polynomial was fitted to the data with a least squares regression. Based on this polynomial, the measurement and linearization uncertainty was analyzed for every TES measurement (see Section \ref{sec:sens}). For $T_{\text{\text{m}}} = \SI{0}{\celsius}$, an example of the errors as a function of the calibration temperature difference $2h$ is presented in Fig. \ref{fig:pler}. The optimal step size is between $2h_{\text{opt}}=\SI{30}{\kelvin}$ and $2h_{\text{opt}} = \SI{50}{\kelvin}$, for the range of mean temperatures $T_{\text{\text{m}}}$ of interest in this study. 

\begin{figure}
\centering
{\includegraphics[]{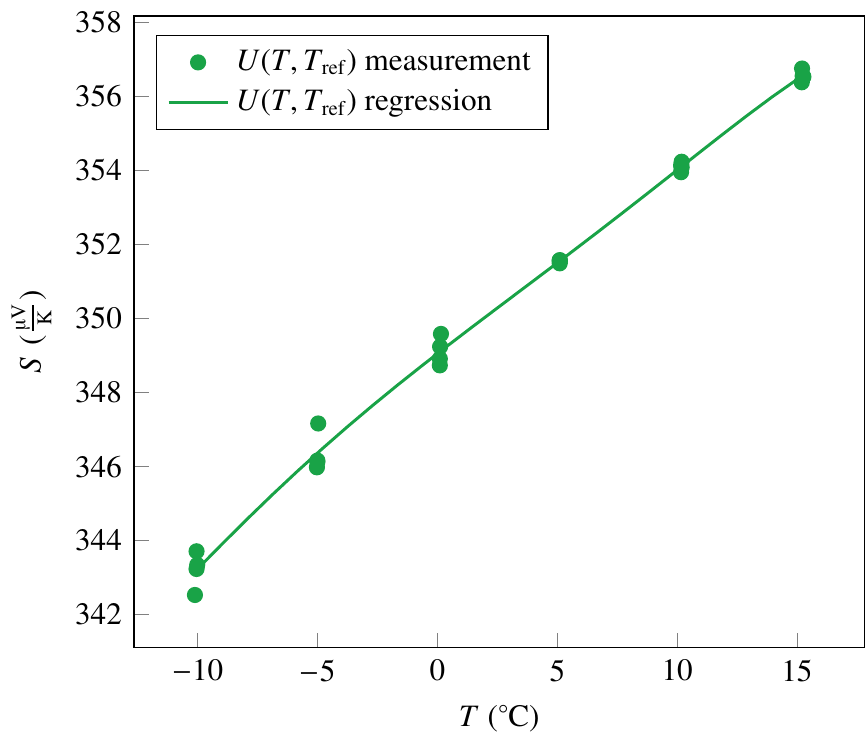}}
\caption{TES $S(T)$ calibration curve. Four measurements were recorded for each of the selected temperatures. 
}
\label{fig:expst}
\end{figure}

The experimental TES data points and the calibration curve $S(T)$ (fitted to the data by a least squares regression with a polynomial of 5th order) are shown in Fig. \ref{fig:expst}. The relative, maximum uncertainty of $S(T)$ is the sum of the relative measurement uncertainty (Eq. (\ref{eqn:me1})) and the relative linearization error $E_{\text{lin}}/S$ (Section \ref{sec:le}). The largest relative, maximum uncertainty is found to be $E_S/S = \SI{2}{\percent}$.

In conjunction with the sensitivity calibration curve and Eq. (\ref{eqn:linvt}), the calibrated thermopile can be employed to measure small differential temperatures
\begin{equation}
\label{eqn:application}
\Delta T = \frac{U}{  {S(T_{\text{\text{m}}})}}  \mathrm{,}
\end{equation}
with $ U$ and $T_{\text{\text{m}}}$ being the measured potential difference and mean temperature respectively. As already discussed in Section \ref{sec:sens}, the TES is relatively robust toward uncertainties in the measurement of the mean temperature. For instance in the worst case scenario, uncertainty in $T_{\text{\text{m}}}$ of $\pm \SI{1}{\kelvin}$ would lead to a possible variation of $S$ by less than $\SI{0.2}{\percent}$. This is negligibly small in comparison to the overall TES uncertainty $E_S/S$ of $\SI{2}{\percent}$. Therefore, measuring $T_{\text{\text{m}}}$ with a level of uncertainty on the order of $\pm \SI{1}{\kelvin}$ suffices in this study.

The resulting maximum uncertainty of a temperature difference measurement is the sum of the relative  sensitivity uncertainty and that due to the multimeter, namely
\begin{align}
\label{eqn:aer}
E_{\Delta T} &= \left(  \frac{E_S}{S}+ \frac{E_U}{U}    \right) \Delta T \nonumber \\
&=\frac{E_S}{S} \Delta T+ \frac{E_U}{S} \mathrm{.}
\end{align}
With the equipment used in this illustrative example, small differential temperatures ($\Delta T \leq \SI{1}{\kelvin}$) can be measured with an uncertainty of $E_{\Delta T} = \pm ( \SI{0.01}{\kelvin}+\SI{2}{\percent} \Delta T)$. The multimeter uncertainty ($E_{U}$) is the limiting factor with respect to differential temperature resolution. It contributes an error on the order of no less than $\SI{0.01}{\kelvin}$ to every measurement. In Table \ref{tab:terror} the uncertainty is shown for various differential temperatures. For example, a temperature difference of $\SI{0.1}{\kelvin}$ can be measured with an uncertainty of $E_{\Delta T} = \pm \SI{0.01}{\kelvin}$.
\begin{table}[]
\centering
\caption{Measurement uncertainty of calibrated thermopile for various differential temperatures.}
 \label{tab:terror}
\begin{tabular}{c | c}
{temperature difference}&{maximum uncertainty}\\  [1mm]
{ $\Delta T$ in $\si{\kelvin}$} & {$E_{\Delta T}$ in $\si{\kelvin}$}\\
\hline
$0.01 $  & $\pm 0.01$ \\[0mm]
$0.05$ & $\pm 0.01$ \\[1mm]
$0.1$ & $\pm 0.01$\\[1mm]
$1.0$ & $\pm 0.03$\\[1mm]
\hline
\end{tabular}
\end{table}

\section{Discussion}
\label{sec:disc}

The uncertainty of small differential temperature measurements with the sensitivity calibrated thermopile can be as low as $\pm \SI{0.01}{\kelvin}$. In comparison, using the calibration Pt100 thermometer ($T \pm \SI{0.3}{\kelvin}$) to measure a differential temperature results in uncertainty of no less than $\pm \SI{0.6}{\kelvin}$. The ratio of the calibration thermometer uncertainty for measuring temperature differences, $2E_{T,\text{cal}}$, and the calibrated thermopile uncertainty for differential temperatures, $E_{\Delta T}$, characterizes the increase in accuracy of the calibration process. The calibration procedure in this study yields a ratio of $2E_{T,\text{cal}}/E_{\Delta T} = 60$.

The sensitivity calibrated thermopile also outperforms other thermopile calibration and signal processing techniques for small temperature differences.  In the method proposed by \citet{Huang1990}, the thermopile is calibrated by comparison to a calibration thermometer with measurement uncertainty of $\SI{0.04}{\kelvin}$. The smallest uncertainty of differential temperature measurements is reported to be $\SI{0.07}{\kelvin}$. Hence, a ratio of $2E_{T,\text{cal}}/E_{\Delta T} = 1.1$ is found for this calibration procedure. In this case, the ratio is based on the most probable error, in contrast to the larger, maximum uncertainty employed in this study. It is therefore somewhat larger than it would be using the maximum uncertainty. An overview of this comparison between the two methods is given in Table \ref{tab:hucomp}.

\begin{table}[h]
\centering
\caption{Comparison of the method proposed by \citet{Huang1990} and the novel approach proposed in this study. Methodological aspects are compared above the line, while the details of the illustrative examples are given below.}
 \label{tab:hucomp}
\begin{tabular}{l | c | c}
& \citet{Huang1990} & proposed approach\\
\hline
calibration coefficient&$U(T)$& $S(T)$  \\[0mm]
$\Delta T$ signal conversion &{midpoint  extrapolation }& linear \\
&$\Delta T = f\left(U(T)_{| T_m \pm \Delta T}\right)$&$\Delta T = U / S(T_m)$\\[1mm]
\hline
instrumentation ${E_{T,cal}}$; ${E_U}$ & $\pm \SI{0.04}{\kelvin}$; $\SI{\pm 5}{\micro \volt}$ & $\pm \SI{0.3}{\kelvin}$; $\pm \SI{3.5}{\micro \volt}$\\[1mm]
max. $\Delta T$ resolution &\SI{0.07}{\kelvin} & $\SI{0.01}{\kelvin}$\\[1mm]
ratio $2E_{T,cal}/E_{\Delta T}$&$1.1$ & $60$\\[1mm]
\hline
\end{tabular}
\end{table}

As is discussed in the previous section, the accuracy of small differential temperature measurements is limited in this study by the uncertainty of the TES $E_S$ and by the uncertainty associated with the employed multimeter $E_U$ (see Eq. (\ref{eqn:aer})). The TES is determined with a relative uncertainty of $E_S/S = \pm 2 \si{\percent}$. Therefore, the corresponding absolute uncertainty of temperature differences scales with the measurand (the measured temperature difference). In contrast to this, the uncertainty due to the multimeter is effectively constant (effectively independent of $\Delta T$) and therefore limits the measurement of small temperature differences. With the multimeter employed in this study the limit is $\pm\SI{0.01}{\kelvin}$. This absolute uncertainty limit is proportional to the multimeter uncertainty. It can be attenuated by employing a more accurate multimeter in the $\Delta T$ measurement process, without the need for renewed calibration.

In general, the same can be achieved by employing thermopiles with a larger number of junctions $N$ (although this was not possible in this study), augmenting $U$. Both of these possibilities lead to reduced relative voltage uncertainty (second term ${E_U}/{U}$ in Eq. (\ref{eqn:aer}) ). This term  is inversely proportional to $N$. For thermopiles with a large number of junctions, this term can become small in comparison to the TES uncertainty (${E_S}/{S}$). The TES uncertainty stems from the calibration procedure and is generally independent of $N$ (given ${E_U}/U<<{E_{2h}}/{2h}$, which is generally the case). Hence, for large thermopile sensitivities, be it by increased multimeter sensitivity or an increase of thermopile sensitivity, the TES uncertainty limits the differential temperature measurement accuracy. Therefore, the method proposed here is particularly effective for devices with high sensitivities.

On the other hand, for the measurement of large temperature differences, the TES uncertainty is the dominant source of uncertainty. For the measurement of temperatures differences larger than the calibration temperature difference $2h$, the calibrated thermopile has lower accuracy than the calibration thermometer. This is no additional limitation to the procedure as it was explicitly developed for small differential temperatures and is limited to these by the linearity assumption associated with Eq. (\ref{eqn:application}).

The novel contribution of the calibration procedure proposed here is the optimized TES measurement. For given apparatus this procedure leads to an optimal temperature difference $2h_{\text{opt}}$ for the differential determination of the TES with minimized uncertainty. This allows the use of larger calibration temperature differences than have been recommended in the past ($\Delta T/ T_{\text{\text{m}}} << 1$) \cite{Zhou2005,Martin2010}, thus reducing the measurement uncertainty. Furthermore, this method also gives an estimate of the magnitude of the linearization error. For instance, herein, the linearization error was less than one quarter the size of the measurement uncertainty. The application of this technique is not limited to the  calibration of thermocouples or thermopiles. It can be applied to optimize the differential method of TES or Seebeck coefficient measurement in general.

To characterize the performance of the optimization procedure on the measurement of TES, we compare the achieved uncertainty using the optimized procedure ($\pm \SI{2}{\%}$) to an equivalent measurement with a commonly employed calibration temperature difference of $2h = \SI{10}{\kelvin}$ \cite{Wold1916,Bidwell1922,Weiss1956,Burkov2001,Zhou2005}. For this temperature difference and with the equipment available in this study, a TES uncertainty of $E^{2h=\SI{10}{\kelvin}}_S = \pm \SI{7}{\%}$ was found. Therefore, in this example, the optimized technique reduces the uncertainty by \SI{70}{\%}.

The reduction of the TES uncertainty also enhances the accuracy of subsequent differential temperature measurements with the calibrated thermopile. For example, a thermopile calibrated using $2h=\SI{10}{\kelvin}$ could measure a temperature difference of $\Delta T=\SI{0.1}{\kelvin}$ with an uncertainty of $\pm \SI{0.02}{\kelvin}$. In comparison, the uncertainty is halved for the thermopile calibrated in this study, which achieves $\Delta T=\SI{0.1}{\kelvin} \pm \SI{0.01}{\kelvin}$. Although this effect may not be crucial for the measurement of relatively large differential temperatures, the use of the optimization procedure is essential for the measurement of heat flux entering a quasi-isothermal heat sink. 

\subsection{Summary of proposed calibration method}
\label{sec:summary}

\begin{figure}[]
\centering
{\includegraphics[]{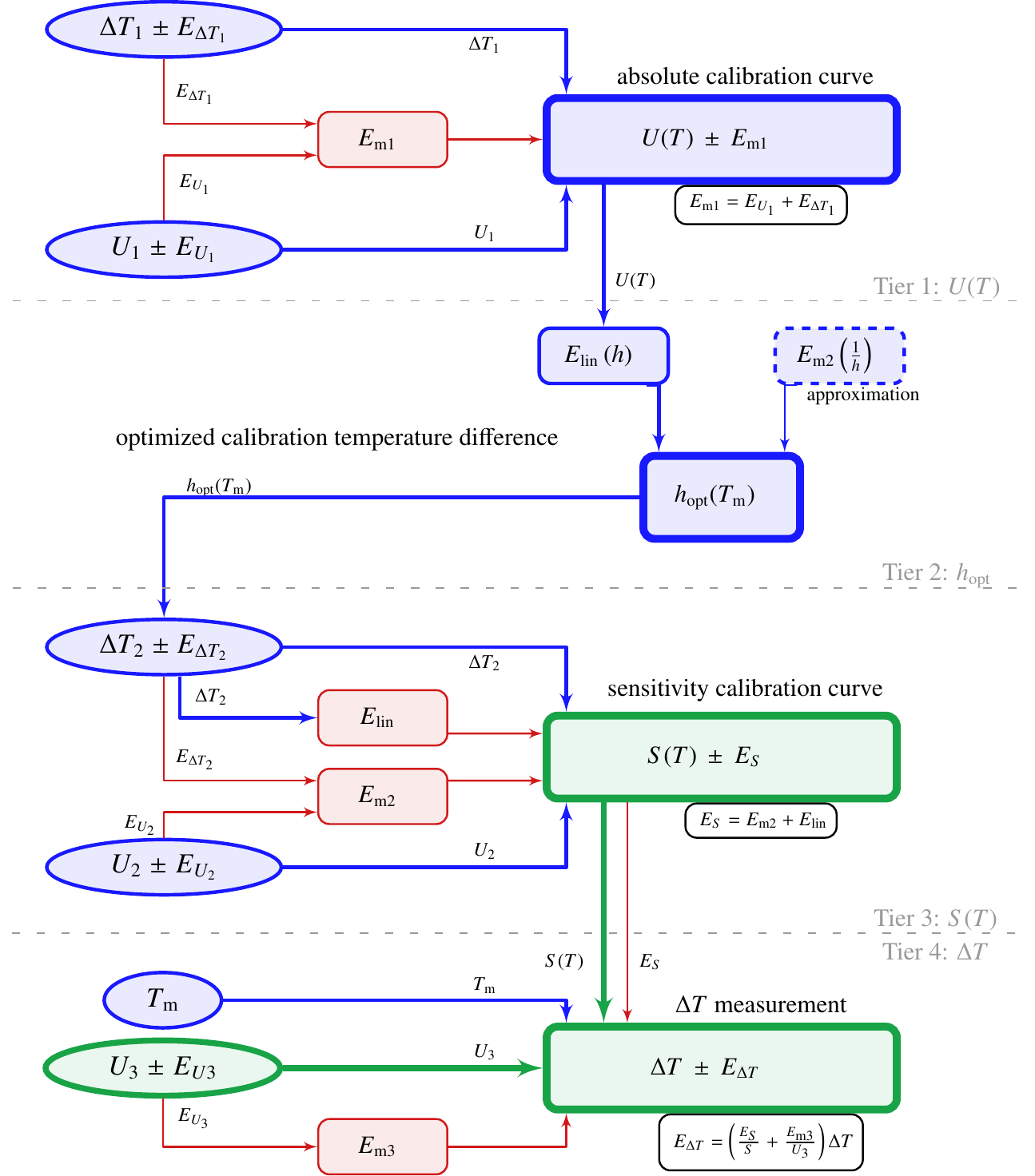}}
\caption{Process diagram of thermopile calibration and differential temperature measurement. Measurements are indicated by ellipses on the left, processed data by rectangles on the right. Data flow is indicated by thick blue and green arrows. Uncertainties and errors are indicated in red boxes in the central column and the propagation thereof in thin red arrows. The thermopile is calibrated in tiers $(1)$ to $(3)$. Tiers $(2)$ and $(3)$ are repeated at intervals over the temperature range of interest. Subsequently the thermopile is employed to measure small differential temperatures, as is shown in tier $(4)$. }

\label{fig:bsb}
\end{figure}

Regardless of the illustrative example given in section \ref{sec:illex} and addressed above, the general sensitivity calibration method to measure the TES can be summarized with the following steps:
\begin{enumerate}
\item[1.] Acquire the standard (absolute) calibration curve $U(T,T_{\text{ref}})$.
\item[2.] Calculate the optimal calibration temperature difference $2h_{\text{opt}}$ based on the minimization of the total TES error $E_S(h) = E_{\text{m}} + E_{\text{lin}}$.
\item[3.] Record the sensitivity calibration curve $S(T)$ using Eq. (\ref{eqn:stm}) and $2h_{\text{opt}}$. Subsequently, analyze the uncertainty $E_S(h)$.
\end{enumerate}
To measure differential temperatures with the calibrated thermopile the following steps need to be performed:
\begin{enumerate}
\item[4.] Measure the mean temperature $T_{\text{\text{m}}}$ to within $\pm \SI{1}{\kelvin}$ and select corresponding sensitivity $S(T_{\text{\text{m}}})$.
\item[5.] Measure the potential difference $U$ over the thermopile.
\item[6.] Calculate the differential temperature with Eq. (\ref{eqn:application}) ($\Delta T = U/S(T_{\text{\text{m}}})$) and the uncertainty $E_{\Delta T}$ (Eq. (\ref{eqn:aer})).
\end{enumerate}
Steps $(1)$ to $(3)$ correspond to tiers $(1)$ to $(3)$ in Fig. \ref{fig:bsb}, while steps $(4)$ to $(6)$ are shown in tier $(4)$. Tiers (2) and (3) are repeated at intervals over the temperature range of interest. 

\section{Conclusions}
\label{sec:conc}
A novel, improved calibration method for small differential temperature sensors based on the Seebeck effect has been developed. To this end, a standard $U(T,T_{\text{ref}})$ calibration curve is recorded to determine an optimal calibration temperature difference $2h_{\text{opt}}$, which minimizes the combined measurement and linearization uncertainties. The overall thermoelectric sensitivity is then measured with the optimized differential (small $\Delta T$) method. Subsequently it is used as the inverse calibration coefficient $1/S$ of a linear calibration relationship $\Delta T = U / S$. 

To demonstrate the proposed method, a thermopile has been fabricated and calibrated specifically for the measurement of small differential temperatures ($\Delta T \leq \SI{1}{\kelvin}$) with an uncertainty of $\pm ( \SI{0.01}{\kelvin}+\SI{2}{\percent} \Delta T)$. In the course of the calibration, the thermoelectric sensitivity has been measured with a maximum relative uncertainty of no more than $E_S/S=\pm \SI{2}{\percent}$. It has been found that in comparison to a measurement with a temperature difference commonly reported in the literature ($2h=\SI{10}{\kelvin}$), the achieved thermoelectric sensitivity uncertainty is $\SI{70}{\%}$ lower. For small differential temperatures the uncertainty of the sensitivity calibrated thermopile is $55$ times lower than that which would be achieved using an alternate signal conversion method and $60$ times lower than employing the calibration thermometer to difference two measured temperatures. 

The optimized calibration method is not only applicable to thermopiles, but also to differential measurement of thermoelectric sensitivity or Seebeck coefficient in general. The highly accurate measurement of differential temperatures is especially of interest in the quantification of heat flux modulations through quasi-isothermal heat sinks.

\section*{Acknowledgements}
\label{sec:ack} The authors gratefully acknowledge the support received from Steffen Hardt. 
 Funding was provided by the German research foundation (DFG grant no. DI 1689/1-1), which is kindly acknowledged.
 
Authors' contributions: 
TP prepared and edited manuscript, conducted measurements, processed data  and developed optimization procedure. 
IN edited manuscript and coordinated the study. 
AM developed optimization procedure and took meaasurements. 
MF and PE fabricated thermopile and performed data analysis. 
MD developed optimization procedure, edited mansucript and guided the study.

\bibliographystyle{elsarticle-num-names}
\bibliography{bib2ETF}

\begin{thebibliography}{36}
\providecommand{\natexlab}[1]{#1}
\providecommand{\url}[1]{\texttt{#1}}
\providecommand{\urlprefix}{URL }
\expandafter\ifx\csname urlstyle\endcsname\relax
  \providecommand{\doi}[1]{doi:\discretionary{}{}{}#1}\else
  \providecommand{\doi}[1]{doi:\discretionary{}{}{}\begingroup
  \urlstyle{rm}\url{#1}\endgroup}\fi
\providecommand{\bibinfo}[2]{#2}

\bibitem[{Hunt(1964)}]{Hunt1964}
\bibinfo{author}{L.~B. Hunt}, \bibinfo{title}{The early history of the
  thermocouple}, \bibinfo{journal}{Platinum Metals Review} \bibinfo{volume}{8}
  (\bibinfo{year}{1964}) \bibinfo{pages}{23--28}.

\bibitem[{Childs et~al.(2000)Childs, Greenwood, and Long}]{Childs2000}
\bibinfo{author}{P.~R.~N. Childs}, \bibinfo{author}{J.~R. Greenwood},
  \bibinfo{author}{C.~A. Long}, \bibinfo{title}{Review of temperature
  measurement}, \bibinfo{journal}{Rev. Sci. Instrum.} \bibinfo{volume}{213}
  (\bibinfo{year}{2000}) \bibinfo{pages}{655--677}.

\bibitem[{van Herwaarden and Sarro(1986)}]{vanHerwaarden1986}
\bibinfo{author}{A.~W. van Herwaarden}, \bibinfo{author}{P.~M. Sarro},
  \bibinfo{title}{Thermal sensors based on the {S}eebeck effect},
  \bibinfo{journal}{Sensor. Actuator.} \bibinfo{volume}{10}
  (\bibinfo{year}{1986}) \bibinfo{pages}{321--346}.

\bibitem[{Bell(2008)}]{Bell2008}
\bibinfo{author}{L.~E. Bell}, \bibinfo{title}{Cooling, heating, generating
  power and recovering waste heat with thermoelectric systems},
  \bibinfo{journal}{Science} \bibinfo{volume}{321} (\bibinfo{year}{2008})
  \bibinfo{pages}{1457--1461}.

\bibitem[{DiSalvo(1999)}]{DiSalvo1999}
\bibinfo{author}{F.~J. DiSalvo}, \bibinfo{title}{Thermoelectric cooling and
  power generation}, \bibinfo{journal}{Science} \bibinfo{volume}{285}
  (\bibinfo{year}{1999}) \bibinfo{pages}{703--706}.

\bibitem[{Riffat and Ma(2003)}]{Riffat2003}
\bibinfo{author}{S.~B. Riffat}, \bibinfo{author}{X.~Ma},
  \bibinfo{title}{Thermoelectrics: a review of present and potential
  applications}, \bibinfo{journal}{Appl. Therm. Eng.} \bibinfo{volume}{23}
  (\bibinfo{year}{2003}) \bibinfo{pages}{913--935}.

\bibitem[{Huang(1990)}]{Huang1990}
\bibinfo{author}{B.~J. Huang}, \bibinfo{title}{A precise measurement of
  temperature difference using thermopiles}, \bibinfo{journal}{Exp. Therm.
  Fluid. Sci.} \bibinfo{volume}{3} (\bibinfo{year}{1990})
  \bibinfo{pages}{265--271}.

\bibitem[{Martin et~al.(2010)Martin, Tritt, and Uher}]{Martin2010}
\bibinfo{author}{J.~Martin}, \bibinfo{author}{T.~Tritt},
  \bibinfo{author}{C.~Uher}, \bibinfo{title}{High temperature {S}eebeck
  coefficient metrology}, \bibinfo{journal}{J. Appl. Phys.}
  \bibinfo{volume}{108} (\bibinfo{year}{2010}) \bibinfo{pages}{121101, 1--12}.

\bibitem[{Armbr\"{u}ster and Kirk(1981)}]{Armbruester1981}
\bibinfo{author}{H.~Armbr\"{u}ster}, \bibinfo{author}{W.~P. Kirk},
  \bibinfo{title}{Millikelvin temperature thermocouple},
  \bibinfo{journal}{Physica B} \bibinfo{volume}{107B} (\bibinfo{year}{1981})
  \bibinfo{pages}{335--336}.

\bibitem[{Maeno et~al.(1983)Maeno, Haucke, and Wheatley}]{Maeno1983}
\bibinfo{author}{Y.~Maeno}, \bibinfo{author}{H.~Haucke},
  \bibinfo{author}{J.~Wheatley}, \bibinfo{title}{Simple differential
  thermometer for low temperatures using a thermocouple with a SQUID detector},
  \bibinfo{journal}{Rev. Sci. Instrum.} \bibinfo{volume}{54}
  (\bibinfo{year}{1983}) \bibinfo{pages}{946--948}.

\bibitem[{Bakker et~al.(2012)Bakker, Flipse, and van Wees}]{Bakker2012}
\bibinfo{author}{F.~L.~. Bakker}, \bibinfo{author}{J.~Flipse},
  \bibinfo{author}{B.~J. van Wees}, \bibinfo{title}{Nanoscale temperature
  sensing using the {S}eebeck effect}, \bibinfo{journal}{J. Appl. Phys.}
  \bibinfo{volume}{111} (\bibinfo{year}{2012}) \bibinfo{pages}{084306, 1--4}.

\bibitem[{Hashemian(2006)}]{Hashemian2006}
\bibinfo{author}{H.~M. Hashemian}, \bibinfo{title}{Maintenance of process
  instrumentation in nuclear power plants}, \bibinfo{publisher}{Springer Berlin
  Heidelberg New York}, \bibinfo{year}{2006}.

\bibitem[{Eccles and Rubart(1973)}]{Eccles1973}
\bibinfo{author}{L.~Eccles}, \bibinfo{author}{W.~Rubart},
  \bibinfo{title}{Differential temperature measurements in engine fluids},
  \bibinfo{journal}{Aircr. Eng. Aerosp. Tec.} \bibinfo{volume}{45 Iss: 6}
  (\bibinfo{year}{1973}) \bibinfo{pages}{27 -- 29}.

\bibitem[{Drnov\v{s}ek et~al.(1998)Drnov\v{s}ek, Pu\v{s}nik, and
  Bojkovski}]{Drnovsek1998}
\bibinfo{author}{J.~Drnov\v{s}ek}, \bibinfo{author}{I.~Pu\v{s}nik},
  \bibinfo{author}{J.~Bojkovski}, \bibinfo{title}{Reduction of uncertainties in
  temperature calibrations by comparison}, \bibinfo{journal}{Meas. Sci.
  Technol.} \bibinfo{volume}{9} (\bibinfo{year}{1998})
  \bibinfo{pages}{1907--1911}.

\bibitem[{Childs et~al.(1999)Childs, Greenwood, and Long}]{Childs1999}
\bibinfo{author}{P.~R.~N. Childs}, \bibinfo{author}{J.~R. Greenwood},
  \bibinfo{author}{C.~A. Long}, \bibinfo{title}{Heat flux measurement
  techniques}, \bibinfo{journal}{P. I. Mech. Eng. C-J. Mec.}
  \bibinfo{volume}{71} (\bibinfo{year}{1999}) \bibinfo{pages}{2959--2975}.

\bibitem[{Langley et~al.(1999)Langley, Barnes, Matijasevic, and
  Gandhi}]{Langley1999}
\bibinfo{author}{L.~W. Langley}, \bibinfo{author}{A.~Barnes},
  \bibinfo{author}{G.~Matijasevic}, \bibinfo{author}{P.~Gandhi},
  \bibinfo{title}{High-sensitivity, surface-attached heat flux sensors},
  \bibinfo{journal}{Microelectr. J.} \bibinfo{volume}{30}
  (\bibinfo{year}{1999}) \bibinfo{pages}{1163--1168}.

\bibitem[{Ewing et~al.(2010)Ewing, Gifford, Hubble, Vlachos, Wicks, and
  T.Diller}]{Ewing2010}
\bibinfo{author}{J.~Ewing}, \bibinfo{author}{A.~Gifford},
  \bibinfo{author}{D.~Hubble}, \bibinfo{author}{P.~Vlachos},
  \bibinfo{author}{A.~Wicks}, \bibinfo{author}{T.Diller}, \bibinfo{title}{A
  direct-measurement thin-film heat flux sensor array}, \bibinfo{journal}{Meas.
  Sci. Technol.} \bibinfo{volume}{21} (\bibinfo{year}{2010})
  \bibinfo{pages}{105201, 1--8}.

\bibitem[{Baughn et~al.(1986)Baughn, Cooper, Iacovides, and
  Jackson}]{Baughn1986}
\bibinfo{author}{J.~W. Baughn}, \bibinfo{author}{D.~Cooper},
  \bibinfo{author}{H.~Iacovides}, \bibinfo{author}{D.~Jackson},
  \bibinfo{title}{Instrument for the measurement of heat flux from a surface
  with uniform temperature}, \bibinfo{journal}{Rev. Sci. Instrum.}
  \bibinfo{volume}{57} (\bibinfo{year}{1986}) \bibinfo{pages}{921--925}.

\bibitem[{Park et~al.(2012)Park, Lee, and Ahn}]{Park2012}
\bibinfo{author}{H.~J. Park}, \bibinfo{author}{D.~H. Lee},
  \bibinfo{author}{S.~W. Ahn}, \bibinfo{title}{An instrument for measuring heat
  flux from an isothermal surface}, \bibinfo{journal}{Exp. Therm. Fluid. Sci.}
  \bibinfo{volume}{37} (\bibinfo{year}{2012}) \bibinfo{pages}{179--183}.

\bibitem[{Koschmieder and Pallas(1974)}]{Koschmieder1974b}
\bibinfo{author}{E.~L. Koschmieder}, \bibinfo{author}{S.~G. Pallas},
  \bibinfo{title}{A sensor for heat transfer measurements},
  \bibinfo{journal}{Rev. Sci. Instrum.}
  \bibinfo{volume}{45}~(\bibinfo{number}{9}) (\bibinfo{year}{1974})
  \bibinfo{pages}{1164--1165}.

\bibitem[{Wold(1916)}]{Wold1916}
\bibinfo{author}{P.~I. Wold}, \bibinfo{title}{The {H}all effect and allied
  phenomena in tellurium}, \bibinfo{journal}{Phys. Rev.} \bibinfo{volume}{7}
  (\bibinfo{year}{1916}) \bibinfo{pages}{169--194}.

\bibitem[{Bidwell(1922)}]{Bidwell1922}
\bibinfo{author}{C.~C. Bidwell}, \bibinfo{title}{Resistance and thermo-electric
  power of metallic {G}ermanium}, \bibinfo{journal}{Phys. Rev.}
  \bibinfo{volume}{19} (\bibinfo{year}{1922}) \bibinfo{pages}{447--455}.

\bibitem[{Weiss(1956)}]{Weiss1956}
\bibinfo{author}{H.~Weiss}, \bibinfo{title}{Bestimmung der effektiven Massen in
  {I}n{S}b und {I}n{A}s aus Messungen der differentiellen Thermospannung},
  \bibinfo{journal}{Zeitschrift f{\"u}r Naturforschung} \bibinfo{volume}{11a}
  (\bibinfo{year}{1956}) \bibinfo{pages}{131--138}.

\bibitem[{Burkov et~al.(2001)Burkov, Heinrich, Konstantinov, Nakama, and
  Yagasaki}]{Burkov2001}
\bibinfo{author}{A.~T. Burkov}, \bibinfo{author}{A.~Heinrich},
  \bibinfo{author}{P.~Konstantinov}, \bibinfo{author}{T.~Nakama},
  \bibinfo{author}{K.~Yagasaki}, \bibinfo{title}{Experimental set-up for
  thermopower and resistivity measurements at $\SI{100}{}$ -
  $\SI{1300}{\kelvin}$}, \bibinfo{journal}{Meas. Sci. Technol.}
  \bibinfo{volume}{12} (\bibinfo{year}{2001}) \bibinfo{pages}{264--272}.

\bibitem[{Zhou and Uher(2005)}]{Zhou2005}
\bibinfo{author}{Z.~Zhou}, \bibinfo{author}{C.~Uher}, \bibinfo{title}{Apparatus
  for {S}eebeck coefficient and electrical resistivity measurements of bulk
  thermoelectric materials at high temperature}, \bibinfo{journal}{Rev. Sci.
  Instrum.} \bibinfo{volume}{76} (\bibinfo{year}{2005})
  \bibinfo{pages}{023901--1 -- 023901--5}.

\bibitem[{Ballestr\'{i}n et~al.(2006)Ballestr\'{i}n, Rodr\'{i}guez-Alonso,
  Rodr\'{i}guez1, Ca{\~{n}}adas, Barbero, Langley, and
  Barnes}]{Ballestrin2006b}
\bibinfo{author}{J.~Ballestr\'{i}n}, \bibinfo{author}{M.~Rodr\'{i}guez-Alonso},
  \bibinfo{author}{J.~Rodr\'{i}guez1}, \bibinfo{author}{I.~Ca{\~{n}}adas},
  \bibinfo{author}{F.~J. Barbero}, \bibinfo{author}{L.~W. Langley},
  \bibinfo{author}{A.~Barnes}, \bibinfo{title}{Calibration of high-heat-flux
  sensors in a solar furnace}, \bibinfo{journal}{Metrologia}
  \bibinfo{volume}{43} (\bibinfo{year}{2006}) \bibinfo{pages}{495--500}.

\bibitem[{Ballestr\'{i}n et~al.(2004)Ballestr\'{i}n, Estrada,
  Rodr\'{i}guez-Alonso, P\'{e}rez-R\'{a}bago, Langley, and
  Barnes}]{Ballestrin2004}
\bibinfo{author}{J.~Ballestr\'{i}n}, \bibinfo{author}{C.~A. Estrada},
  \bibinfo{author}{M.~Rodr\'{i}guez-Alonso},
  \bibinfo{author}{C.~P\'{e}rez-R\'{a}bago}, \bibinfo{author}{L.~W. Langley},
  \bibinfo{author}{A.~Barnes}, \bibinfo{title}{High-heat-flux sensor
  calibration using calorimetry}, \bibinfo{journal}{Metrologica}
  \bibinfo{volume}{41} (\bibinfo{year}{2004}) \bibinfo{pages}{314--318}.

\bibitem[{Bernhard(2004)}]{Bernhard2004}
\bibinfo{editor}{F.~Bernhard} (Ed.), \bibinfo{title}{Technische
  Temperaturmessung}, \bibinfo{publisher}{Springer-Verlag Berlin Heidelberg New
  Yok}, \bibinfo{year}{2004}.

\bibitem[{{EN 60584-1}(1995)}]{DIN60584}
\bibinfo{author}{{EN 60584-1}}, \bibinfo{title}{Thermopaare Teil 1: Grundwerte
  der Thermospannung (EN 60584-1:1995)}, \bibinfo{year}{1995}.

\bibitem[{{EURAMET cg-8}(2011)}]{EURAMET}
\bibinfo{author}{{EURAMET cg-8}}, \bibinfo{title}{Calibration of Thermocouples,
  EURAMET cg-8 Version 2.1}, \bibinfo{year}{2011}.

\bibitem[{Bentley(1998)}]{Bentley1998}
\bibinfo{author}{R.~E. Bentley}, \bibinfo{title}{Handbook of temperature
  measurement vol. 3: theory and practice of thermoelectric thermometry},
  \bibinfo{publisher}{Springer}, \bibinfo{year}{1998}.

\bibitem[{Drebuschak(2009)}]{Drebuschak2009}
\bibinfo{author}{V.~A. Drebuschak}, \bibinfo{title}{Universality of the emf of
  thermocouples}, \bibinfo{journal}{Thermochimica Acta} \bibinfo{volume}{496}
  (\bibinfo{year}{2009}) \bibinfo{pages}{50--53}.

\bibitem[{Wood et~al.(1988)Wood, Chmieleski, and Zoltan}]{Wood1988}
\bibinfo{author}{C.~Wood}, \bibinfo{author}{A.~Chmieleski},
  \bibinfo{author}{D.~Zoltan}, \bibinfo{title}{Measurement of {S}eebeck
  coefficient using a large thermal gradient}, \bibinfo{journal}{Rev. Sci.
  Instrum.} \bibinfo{volume}{59} (\bibinfo{year}{1988})
  \bibinfo{pages}{951--954}.

\bibitem[{Pinisetty et~al.(2011)Pinisetty, Haldolaarachige, Young, and
  Devireddy}]{Pinisetty2011}
\bibinfo{author}{D.~Pinisetty}, \bibinfo{author}{N.~Haldolaarachige},
  \bibinfo{author}{D.~Young}, \bibinfo{author}{R.~V. Devireddy},
  \bibinfo{title}{A novel experimental device for {S}eebeck coefficient
  measurements of bulk materials, thin films, and nanowire composites},
  \bibinfo{journal}{J. Nanotechnol. Eng. Med.} \bibinfo{volume}{2}
  (\bibinfo{year}{2011}) \bibinfo{pages}{011006--1 -- 011006--5}.

\bibitem[{Middleton and Scanlon(1956)}]{Middleton1953}
\bibinfo{author}{A.~E. Middleton}, \bibinfo{author}{W.~W. Scanlon},
  \bibinfo{title}{Measurement of the thermoelectric power of {G}ermanium at
  temperatures above $\SI{78}{\degree \kelvin}$}, \bibinfo{journal}{Zeitschrift
  f{\"u}r Naturforschung} \bibinfo{volume}{11a} (\bibinfo{year}{1956})
  \bibinfo{pages}{131--138}.

\bibitem[{Sch\"{a}fer(2006)}]{Schafer2006}
\bibinfo{author}{M.~Sch\"{a}fer}, \bibinfo{title}{Computational engineering -
  introduction to numerical methods}, \bibinfo{publisher}{Springer-Verlag
  Berlin Heidelberg}, \bibinfo{year}{2006}.

\end{thebibliography}

\end{document}